\documentstyle[epsfig,12pt]{article}
\begin{document} 
 \newcommand \be {\begin{equation}}
\newcommand \ee {\end{equation}}
 \newcommand \ba {\begin{eqnarray}}
\newcommand \ea {\end{eqnarray}}

\newcommand{\siml}{\stackrel{<}{\sim}}

\title{Elements for a Theory of Financial Risks}
\author{J.-Ph. Bouchaud}
\date{Service de Physique de l'\'Etat Condens\'e,
 Centre d'\'etudes de Saclay, \\ Orme des Merisiers, 
91191 Gif-sur-Yvette Cedex, France \\Science \& Finance, 109-111 Rue Victor
Hugo, \\
92 323 Levallois Cedex. }

\maketitle
\begin{abstract}
Estimating and controlling large risks has become one of the
main concern of
financial institutions. This requires the development of adequate statistical
models and theoretical 
tools (which go beyond the traditionnal theories based on Gaussian
statistics), and their practical implementation. Here we describe three
interrelated aspects of this program: we first give a brief survey of the
peculiar statistical properties of the empirical price fluctuations. We then
review how an option pricing theory consistent with these statistical features
can be constructed, and compared with real market prices for options. We
finally argue that a true `microscopic' theory of price fluctuations (rather
than a statistical model) would be most valuable for risk assessment. A simple
Langevin-like equation is proposed, as a possible step in this direction. 
\end{abstract}

\section{Introduction}

The efficiency of the theoretical tools devised to address the problems of
risk control, portfolio selection and derivative pricing strongly depends on 
the adequacy of the stochastic model chosen to describe the market
fluctuations. Historically,
the idea that price changes could be modelled as a Brownian motion dates
back to Bachelier \cite{Bachelier}. This hypothesis,
 or some of its variants (such as the Geometrical Brownian  motion, where the
log of the price is a Brownian motion) is at the root of most of the modern
results of
 mathematical finance, with Markowitz portfolio analysis, the Capital Asset
Pricing Model ({\sc capm}) \cite{RefGen} and the  Black-Scholes formula
\cite{BS}  standing
out as
paradigms. The reason for success is mainly due to the impressive
mathematical
and probabilistic apparatus available to deal with Brownian
 motion  problems, in particular  Ito's stochastic calculus.

An important justification of the Brownian motion description lies in the
Central Limit Theorem ({\sc clt}), stating
 that the sum of $N$ identically distributed, weakly dependent random
changes is, for large $N$, a
Gaussian variable. In physics or in finance, the number
of elementary changes observed during a time interval $t$ is given by
 $N={t \over \tau^*}$ where $\tau^*$ is an elementary correlation time,
below which changes of velocity (for the case of a Brownian particle)
 or changes of `trend' (in the case of the stock prices) cannot occur.
The use of the {\sc clt} to substantiate
the use of Gaussian statistics in any caserequires that $t \gg \tau^*$.
 In financial markets, $\tau^*$ turns out to be of the order of several
minutes,
 which is not that small compared to the relevant time scales (days),
 in particular when one has to
worry about the {\it tails} of the distribution (corresponding to large
 shocks) which sometimes disappear only very slowly. 

The fact that the tails of the distribution of returns are much `fatter' than 
predicted by the gaussian is well known, in particular since the seminal work
of Mandelbrot \cite{Man}, where the idea that price changes are still
independent, but distributed according to a L\'evy stable law was 
first proposed. This model however fails in two respects. First, the tails of
the distribution of returns is now much {\it overestimated}; in particular,
the variance appears to be well defined in most financial markets, while it is
infinite for L\'evy distributions. Second, and perhaps more importantly in
view of its application to option markets, the amplitude of the fluctuations
(measured, say, as the local variance) appears to be itself a randomly
fluctuating variable with rather long range
correlations. 

The aim of this paper is to provide a short survey of the most prominent 
statistical properties of the fluctuations of rather liquid markets, which are
characterized by what one could call `moderate' fluctuations (for a review,
see
e.g. \cite{Olsen}). Less liquid 
markets sometimes behave rather differently and more `extreme' fluctuations
can be observed. We shall then present a theory for option pricing
and hedging in the general case where the underlying stock fluctuations are
not gaussian. In this case, perfect hedging is in general
impossible, but optimal strategies can be found (analytically or numerically)
and the associated residual risk can be estimated. We show that the volatility
`smile' observed on option markets can be understood using a cumulant
expansion, and discuss the idea of an implied `kurtosis', which is (on liquid
markets) very close to the actual  (maturity dependent) kurtosis of the
historical data.
Finally, we briefly discuss a simple `Langevin' approach to market
fluctuations, which aims at describing, in a coarse-grained way, the feedback
effects between the agents behaviour and the price fluctuations.
Interestingly, a natural distinction appears between a normal regime, where
price fluctuations are moderate, 
from a `crash' regime, where panic effects are self-reinforcing and
leads to a price collapse. 

\section{A Short Survey of Empirical Data}
\label{Empirical}
\subsection{Linear correlations}

We shall denote in the following the value of the stock (or any other asset)
at time $t$ as $x(t)$,
and the variation of the stock on a given time interval
$\tau$ as
$\delta_\tau x(t) = x(t+\tau)-x(t)$. The time delay $\tau$ can be as small as
a few seconds in actively traded markets. However, on these short time scales,
the fluctuations cannot be
considered to be independent. For example, the second order correlation
function defined as
\footnote{In principle, the average value of $\delta_\tau x(t)$ should be
removed, but
this leads to insignificant corrections on the time scales considered. For the
same reason, we
neglect the difference between the variation of the price and the more often
studied 
variation of the log of the price.}:
\be
C(\Delta t) = \frac{\langle \delta_\tau x(t+\Delta t) \delta_\tau x(t)
\rangle}
{\langle  \delta_\tau x(t)^2 \rangle}
\ee
is significantly non-zero up to $\Delta t =
15$ minutes, as shown in Fig. 1.

\begin{figure}
\centerline{\hbox{\epsfig{figure=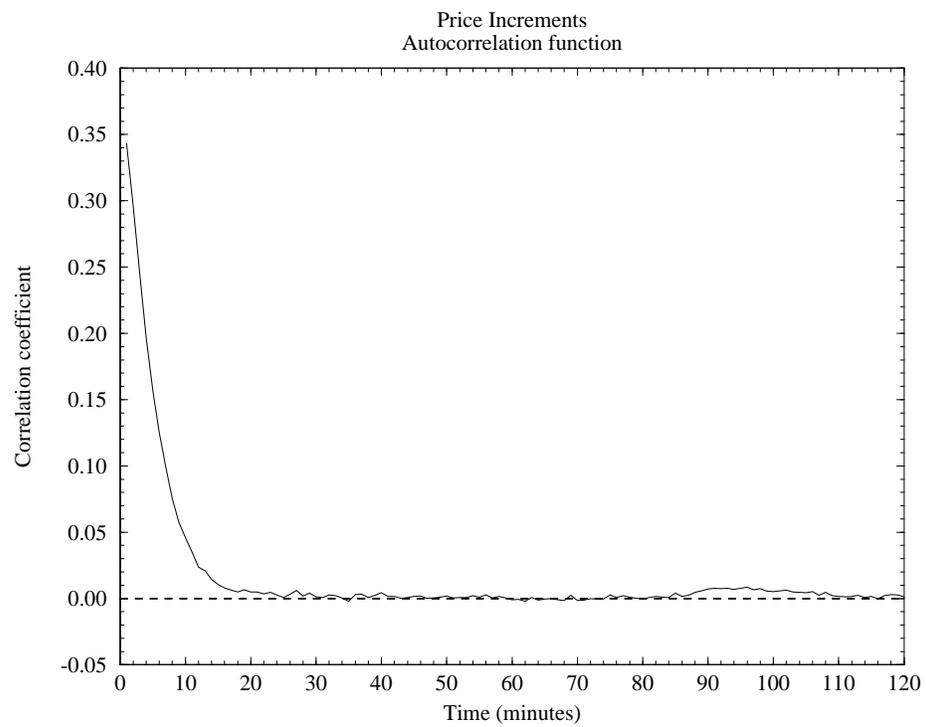,width=10cm,angle=270}}}
\vskip 0.8cm
\caption{\small{Correlation function $C(\Delta t)$ of the one minute price
increments of the S\&P 500, for the period 1983-1996,
which decays to zero on the scale of $\tau^* \simeq 15$ minutes. Note that
$\tau^*$ has drifted to smaller values with time.}}
\label{fig1}
\end{figure}

Therefore, it is certainly inadequate to think of the price process as a
`martingale' for 
small time increments. This however does not necessarily mean that there are
arbitrage
opportunities: one can easily see that even very small transaction costs
prevent the use
of these short time correlations, which would imply a very high (and thus very
costly) trading frequency \cite{BP}. For time delays $\Delta t$ longer than a
certain
$\tau^*$, say one hour, the autocorrelation of price increments is very nearly
zero; correspondingly, the variance of price
increments grow linearly with $\Delta t$ for $\Delta t > \tau^*$.

\subsection{Distribution of elementary increments}

The simplest idea is thus that the increments \footnote{In the following,
we shall simplify the notation and use $\delta(t)\equiv \delta_{\tau^*} x(t)$
for the increments on the time scale $\tau^*$.}
$\delta(t)$ are independent identically distributed ({\sc iid}) random
variables. In this
case, the knowledge of the distribution density $P^*$ of 
$\delta$ would suffice to reconstruct the distribution of increments on any
time 
delay $\Delta t$ larger than $\tau^*$, through a simple convolution. The shape
of $P^*$ is strongly non gaussian: an estimate of its kurtosis
$\kappa_{\tau^*}$ on a historical basis leads to numbers on the order of $20$
(again on liquid markets). As shown in Fig. 2 on the example of the British
Pound/U.S. \$ time series, the tails of $P^*$ decay as an exponential:
\be
P(\delta) \propto \exp\left(-\lambda_\pm |\delta|\right) \qquad \delta \to \pm
\label{exp}\infty
\ee
(although several authors have reported a somewhat slower decay, as a power
law $|\delta|^{-1-\nu}$ with a rather large exponent $\nu \simeq 3$--$4$
\cite{frechet}, or possibly a `stretched exponential' \cite{sornette}). Such
slowly decaying tails survive upon
convolution, and are thus particularly relevant for extreme risks forecasts
\cite{Profiler}. 

\vskip 0.2cm
\begin{figure}
\centerline{\hbox{\epsfig{figure=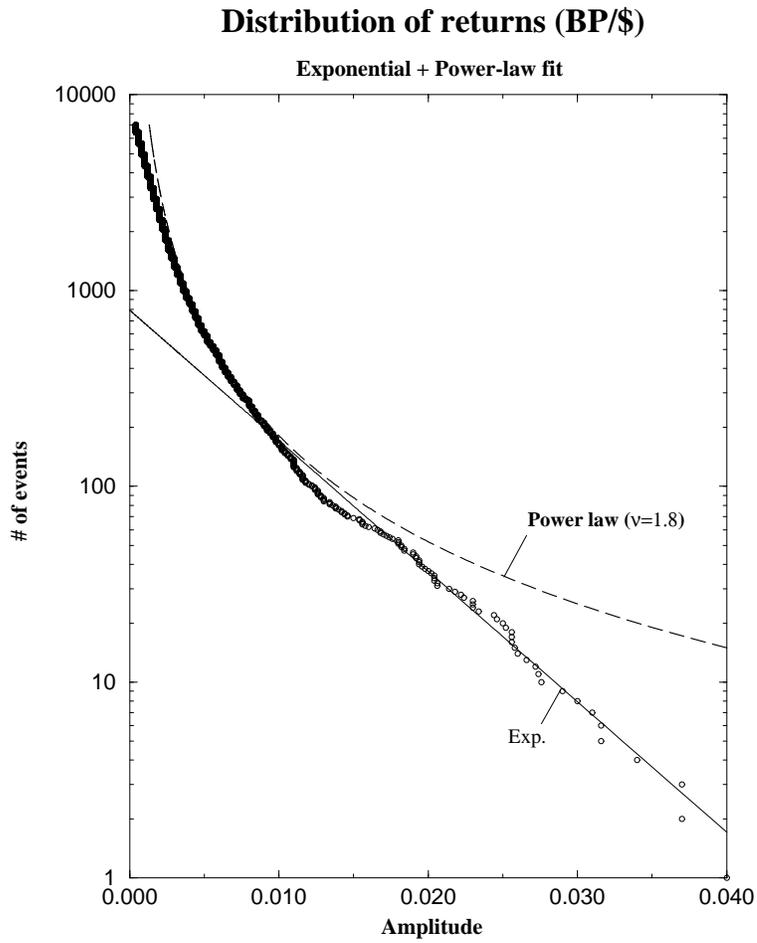,width=10cm}}}
\vskip 0.8cm
\caption{\small{Distribution of the British Pound/US \$ negative increments
for $\tau^*=30$ mimutes, in the period 1991-1995. The `near-tail' of the
distribution can be fitted by a power-law, which clearly overestimates the
`far-tail'. The latter is well represented by an exponential fall-off, 
Eq. (\protect\ref{exp}): note the linear-log scale}} \label{fig2}
\end{figure}

A reasonable fit of $P^*$ on most markets can be achieved using a symmetrical
`truncated' L\'evy distribution\cite{Stanley,BP}, defined in Fourier space as:
\be
\log P^*(z) = \frac{A}{\cos \pi \mu/2}\left[{\lambda^\mu}
- (\lambda^2+z^2)^{\mu/2} f(z) \right];\quad f(z)=\cos \left(\mu \arctan
\frac{|z|}{\lambda}\right)
\ee
where $A$ the scale factor, $\mu$ is the `tail exponent', which is
found to
be  close to $1.5$ for all markets, and $\lambda$ describes the exponential
fall of the far-tails.
Note that when
$\mu=2$, one recovers the gaussian distribution, while for $\lambda \to 0$,
one finds the
stable L\'evy distributions proposed by Mandelbrot \cite{Man}, with tails
which decay as
$|\delta|^{-1-\mu}$.

\subsection{Anomalous decay of the kurtosis and volatility persistence}

The $N^{th}$ autoconvolution of $P^*$, where $N=\Delta t/\tau^*$ is simply
obtained by scaling the parameter $A$
to $NA$. With no
further ajustable parameters, this leads to a fair representation of the
historical distribution of
$\delta_{\Delta t} x$ \cite{BP,comment}. However, some systematic differences
(in particular in
the tails of the distribution) show up, which reveal the inadequacy of such a
simple {\sc iid} hypothesis. For example, it is easy to show that under such
an hypothesis, the kurtosis of the increments on scale $\Delta t=N\tau^*$
should be a factor $N$ smaller than the kurtosis on scale $\tau^*$.
Empirically, however, one finds a much slower decay, as $\kappa_N =
\kappa_1/N^\lambda$, with $\lambda$ in the range $0.3-0.6$ (see Fig. 3). This
can be
related to another observation, which was made many times in the literature
\cite{Olsen,archeffects}:
higher order correlation functions of the price increments, such as:
\be
C_2(m) = \frac{\langle \delta^2(t+m \tau^*) \delta^2(t)
\rangle}
{\langle  \delta^2(t) \rangle^2} -1 
\ee
decay only very slowly with time. A simple fit of this decay is again of the
power law type \cite{Olsen,cizeau,arneodo}:
\be
C_2(m) \simeq \frac{C_{0}}{m^{\lambda'}}\label{power}
\ee
with a exponent $\lambda'$ in the same range as $\lambda$. A simple way to
rationalize these findings is to assume that while the {\it sign} of the price
increment is
completely decorrelated as soon as $\Delta t > \tau^*$, the amplitude of the
increment (which is a measure of the market activity) is correlated in time.
Bursts of market activity, related to external news or crisis, can easily
persist for several days, sometimes months -- this explains why $C_2(m)$
decays rather slowly.
We shall thus assume that the increment $\delta$ can be written as:
\be
\delta(t_i) = \gamma(t_i) \times \epsilon(t_i) \qquad t_i = i \tau^*
\label{randomscale}
\ee
where the scale $\gamma > 0$ measures the amplitude of the increment,
which can be thought of as the local volatility of the market. The random
variable $\epsilon$ is 
short range correlated (over time $\tau^*$) of mean zero and
variance unity (and independent from $\gamma$). Note that $\epsilon$ is not
necessarily Gaussian, as assumed in {\sc arch}-like models \cite{archeffects}.
It is then rather easy to show
that the 
kurtosis of the increments on scale $\Delta t=N\tau^*$ can be expressed as
\cite{Potters,BP}:
\be
\kappa_N = \frac{1}{N} \left[\kappa_0 + (3+\kappa_0)C_2(0) + 6 \sum_{m=1}^N
(1-\frac{m}{N}) C_2(m) \right]\label{gm} 
\ee
where $\kappa_0$ is the kurtosis of the random variable $\epsilon$. Note that
even if $\kappa_0=0$, the kurtosis of $P^*$ is non zero due to the randomly
fluctuating
scale parameter $\gamma$. If one furthermore assumes that $C_2$ decays as a
power-law (Eq. \ref{power}), then from Eq. (\ref{gm}), $\kappa_N$ decays for 
large $N$ as $N^{-\lambda}$ with $\lambda=\lambda'$.

\subsection{Apparent (multi)-scaling behaviour}

Hence, the fact that the scale $\gamma$ of the random increments has long
range temporal correlations induces an anomalously slow convergence of the sum
$x({N\tau^*})-x(0)=\sum_{i=1}^N \delta(t_i)$ towards the gaussian
distribution. One should note that this can induce {\it apparent} scaling
behaviour on restricted time intervals. For example, a numerical simulation of
the random scale model (\ref{randomscale}) with a slowly decaying $C_2(m)$ can
be analyzed in scaling terms \cite{unpub}, i.e. fitting the moments
of $x({N\tau^*})-x(0)$ as power laws:
\be
\langle |x({N\tau^*})-x(0)|^q \rangle \propto N^{\zeta_q}
\ee
This actually works quite well, and leads to a family of exponents $\zeta_q$
which
deviates from the theoretical straight line $\zeta_q = q/2$ which holds in the
limit
$N \to \infty$, provided all the moments of $\gamma$ exist. For $q=2$, the
relation $\zeta_2=1$ holds because the linear correlation function $C(\Delta
t)$ is zero for $\Delta t > \tau^*$. For $q=4$ one finds, using the definition
of the kurtosis:
\be
\langle [x({N\tau^*})-x(0)]^4 \rangle = (3 + \kappa_N) \langle
[x({N\tau^*})-x(0)]^2 \rangle^2 \propto 3 N^2+ \kappa_1 N^{2-\lambda}
\ee
which is thus the sum of two power laws. This can be however fitted with a
unique `effective' exponent $\zeta_4$, which can be substantially below the
asymptotic value $\zeta_4^{\infty}=2$ since $N^{-\lambda}$ is not very small
in practice. This argument holds true for higher 
moments for which $\zeta_q < q/2$; the difference between $\zeta_q$ and $q/2$ 
actually grows with $q$ (for a fixed value of $N$). We believe that this might
be the reason for the `multifractal' scaling recently reported in the
literature 
\cite{multifractal,multifM}.

\subsection{Time reversal symmetry}

The last empirical fact which we would like to comment on is the question of
time 
reversal symmetry. Is it possible to detect the direction of time in a
financial 
time series ? The answer to this question would be no for a simple Brownian
motion, 
for example, but also for a much wider class of processes, such as the
multifractal
time construction proposed in \cite{multifM}. Correlation functions sensitive
to the arrow of time have been proposed by 
Pomeau \cite{Pomeau}. One example is:
\be
C_{T}(\Delta t)=\langle x(t)\left[x(t+\Delta t)-x(t+2\Delta
t)\right]x(t+3\Delta t) \rangle
\ee
which is non zero if the time triplets $t,t+\Delta t,t+3\Delta t$ and 
$t,t+2 \Delta t,t+3\Delta t$ cannot be distinguished statistically. For the
price series itself,
the above correlation function is zero within error bars. But when one studies
the
`volatility' process $\gamma$, then this skew correlation function is
distinctively 
non zero, 
and reaches a maximum (in the case of the S\&P 500) for $\Delta t \simeq$ 1
month.  
This shows that the volatility time series is {\it not} invariant under time
reversal. 
A similar conclusion is also reported in \cite{Dacorogna2,arneodo}, where the
authors 
observe that a high `coarse-grained' volatility in the past increases in a
causal way 
today's `fine-grained' volatility. This is not unreasonable, as one feels
intuitively 
that an anomalously large change of the close to close price over -- say --
the previous week triggers 
more intraday activity the following week. In any case, we feel that this
absence of time reversal symmetry is an important (albeit less frequently
discussed) stylized fact of financial time series.

\section{Implications for Option Pricing}
\subsection{General framework}

We now turn to the problem of option pricing and hedging when the statistics 
for price increments have the non-Gaussian properties discussed above. The 
distinctive feature of the continuous time random walk model usually
considered
in the theory of option pricing is the possibility of perfect hedging
\cite{Hull}, that
is, 
a complete elimination of the risk associated to option trading \cite{BS}. 
This property
however no longer holds for more realistic models \cite{BS94}. 

Let us write down the global wealth balance
$\Delta W|_0^T$ associated with the writing of a `call' option of maturity $T$
and exercice price $x_s$ \cite{BP}:
\ba
\Delta W|_0^T &=&  {\cal C}(x_0,x_s,T)\exp(rT)  -
\max(x(T)-x_s,0) \\
 &+& \sum_i  \ \phi(x,t_i)\exp(r(T-t_i)) [\delta_i -rx(t_i)\tau],
\ea
where ${\cal C}(x_0,x_s,T)$ is the price of the call, $x_0=x(t=0)$ and
$\phi(x,t)$ the
trading strategy, i.e. the number of stocks per option in the portfolio of the
option writer. Finally, $r$ the (constant) interest rate.
The second term defines the option contract: the profit of the buyer of the
option
is equal to $x_s-x(T)$ if $x(T) > x_s$ (i.e. if the option is exercised) and
zero otherwise: the option
is an insurance contract which guarantees to its owner a maximum price for
acquiring a certain stock at time $T$. Conversely, a `put' option would
guarantee
a certain minimum price for the stock held by the owner of the option.

A natural procedure to fix the price of the option ${\cal C}(x_0,x_s,T)$ and
the optimal
strategy $\phi^*(x,t)$ was proposed independently in \cite{Schweitzer,BS94}
and
further discussed in \cite{BP,AS,Matacz}. It consists in imposing a 
{\it fair
game condition}, i.e.:
\be
\langle \Delta W|_0^T[\phi] \rangle = 0 \label{13}
\ee
and a {\it risk minimisation condition}:
\be
\left.\frac{\delta \langle \Delta W|_0^T[\phi]^2 \rangle}{\delta
\phi(x,t)}\right|_{\phi^*} = 0\label{min}
\ee
Here, we assume that the variance of the wealth variation is a relevant
measure of the risk. However, other measures are possible, in particular the
`Value-at-Risk', 
which is directly related to the weight contained in the negative tails of the
distribution of $\Delta W|_0^T$. 

The notation $\langle ... \rangle$ in Eqs. (\ref{13},\ref{min}) means that
one averages over the probability of the different trajectories. The explicit
solution of Eqs. (\ref{13},\ref{min}) for a general uncorrelated process (i.e.
$\langle
\delta_i \delta_j \rangle =0$ for $i \neq j$) is relatively easy to write if
the average bias $\langle \delta \rangle$ and the interest rate $r$ are
negligible\footnote{For the general case, see
\cite{BP,AS}},
which is the case for short maturities $T$. In this case, one finds
\cite{BS94}:
\be
{\cal
C}(x_0,x_s,T) = \int_{x_s}^\infty dx' \ (x'-x_s)
P(x',T|x_0,0)\label{price}
\ee

\be
\phi^*(x,t) =  \int_{x_s}^\infty dx'
\langle\delta\rangle_{(x,t)\longrightarrow
(x',T)}
{(x'-x_s) \over {\sigma^2}(x,t)} P(x',T|x,t) 
\ee
where $\sigma^2(x,t)=\langle \delta^2 \rangle|_{x,t}$ is the `local
volatility' -- which may
depend on $x,t$ -- and $\langle \delta \rangle _{(x,t)\longrightarrow
(x',T)}$
is
the mean
instantaneous
increment conditioned to the initial condition $(x,t)$ and
 a final condition $(x',T)$. The {\it minimal} residual risk, defined as
${\cal R}^*=\langle \Delta W|_0^T[\phi^*]^2 \rangle$ is in general strictly
positive (and in practice rather large),
except for Gaussian fluctuations {\it in the continuous limit}, where the 
residual risk is strictly zero! In this limit, the above equations
(\ref{13},\ref{min}) actually exactly lead to the celebrated Black-Scholes
option
pricing formula. In particular, one can check that $\phi^*$ is related to
${\cal C}$ through: $\phi^*=\partial{\cal
C}(x_0,x_s,T)/{\partial x_0}$.

\subsection{Cumulant expansion and volatility smile}

In the case where the market fluctuations are moderately non-Gaussian, one
might expect that a {\it cumulant expansion} around the Black-Scholes formula
leads to 
interesting results. This cumulant expansion has been worked out in general in
\cite{BP}. If one only retains the leading order correction which is 
(for
symmetric fluctuations) proportional to the kurtosis, one finds that the 
price of options ${\cal C}(x_0,x_s,T)$ can be written as a Black-Scholes
formula, but with a modified value of the volatility $\sigma$, which
becomes price and maturity dependent \cite{Potters}:
\be 
\sigma_{imp.}(x_s,T) = \sigma
\left[1+ \frac{\kappa_T 
}{24}
\left(\frac{(x_s-x_0)^2}{\sigma^2 T}- 1
\right)\right] 
\label{smile}
\ee 
The volatility $\sigma_{imp.}$ is called the implied volatility by the market
operators, who use the standard Black-Scholes formula to price options, but
with a
value of the volatility which they estimate intuitively, and which turns out
to depend on
the exercice price in a roughly parabolic manner, as indeed suggested by Eq.
(\ref{smile}).

\begin{figure}
\centerline{\hbox{\epsfig{figure=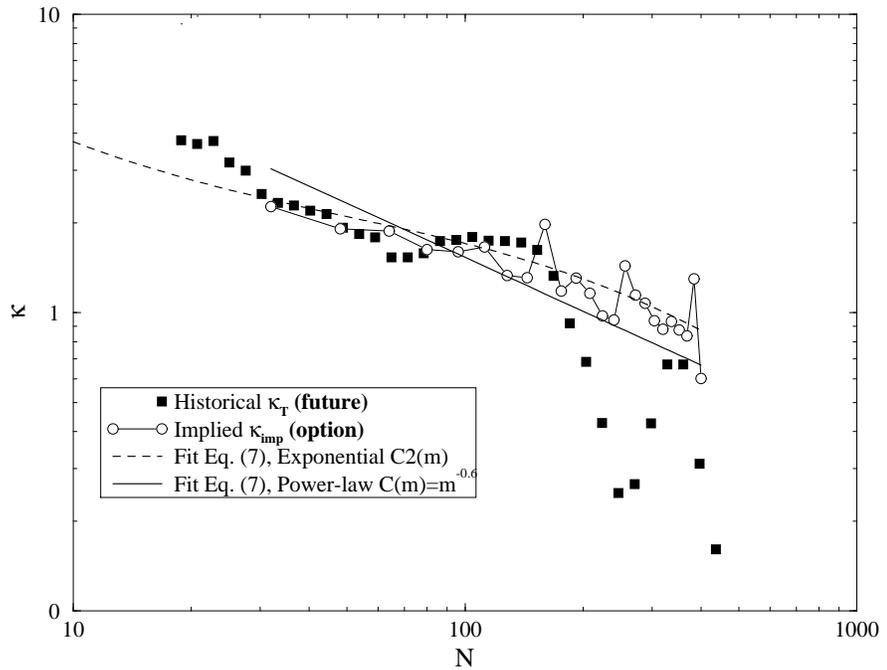,width=10cm,angle=270}}}
\vskip 0.3cm
\caption{\small{Plot (in log-log coordinates) of the average implied kurtosis
$\kappa_{imp}$ (determined by fitting the implied volatility for a fixed
maturity by a parabola)
and of the empirical kurtosis $\kappa_N$ (determined directly
from the
historical movements of the {\sc bund} contract), as a function of the
reduced time
scale $N=T/\tau$, $\tau = 30$ minutes. All
transactions of options on the
{\sc bund} future from 1993 to 1995 were analyzed
along with 5 minute tick data of the
{\sc bund} future for the same period. We show for comparison a fit with 
$\kappa_N \simeq N^{-0.6}$ (dark line). A fit with an exponentially decaying
$C_2(m)$ is however also acceptable (dotted line).
}}
\label{fig3}
\end{figure}

This is the so-called `volatility smile'. Eq. (\ref{smile})
furthermore shows that the curvature of the smile is directly related to the
kurtosis of the underlying statistical process on the scale of the maturity
$T=N\tau^*$. We have tested this prediction by directly comparing the `implied
kurtosis', obtained by extracting from real option prices (on the {\sc bund}
market) the volatility $\sigma$ (which turns out to be highly correlated with
a short time filter of the historical volatility), and the curvature of the
implied volatility smile, to the historical value of the kurtosis $\kappa_N$
discussed above. The result is plotted in Fig. 3, with no further ajustable
parameter. The remarkable agreement between the implied and historical
kurtosis, and the fact that they evolve similarly with maturity, shows 
that the market as a whole is able to correct (by trial and errors) the
inadequacies of the Black-Scholes formula, and to encode in a satisfactory way
both the fact that
the distribution has a positive kurtosis, and that this kurtosis decays in an
anomalous fashion due to volatility persistence effects. However, the real
risks associated with option trading are, at present, not satisfactorily
estimated. In particular, most risk control softwares dealing with option
books are based on a Gaussian description of the fluctuations.

\section{A Langevin Approach to Price Fluctuations: Feedback Effects}

In the author's opinion, there is as yet no convincing `microscopic'
model which explains the distinctive statistical features of price
fluctuations which were summarized in section \ref{Empirical}, although many
proposals have been put forward \cite{models1,models2}. In the spirit of
statistical physics, it might be
possible to describe market dynamics at a level of description which is 
intermediate between the macroeconomic level
which is that of the market equilibrium models \cite{models1}
and the individual agent level
which is that of the market microstructure theory.
Our basic idea is that although the modelling of each individual participants
(`agents') is impossible in quantitative terms, the collective behavior of
the market and its impact on the price in particular can be represented
in statistical terms  by a few number of terms
in a (stochastic) dynamical equation. Our approach is in the spirit of many
phenomenological, `Landau-like' approaches to physical phenomena 
\cite{hwakardar}. We have thus proposed a description of the dynamics of
speculative markets with
a simple Langevin equation \cite{Langevin}. 
This equation encapsulates what we believe to be 
some essential ingredients; in
particular, the feedback of the price fluctuations on the behaviour of the
market participants. The construction of a convincing model for large
fluctuations is important for risk control, since empirical data may be
unreliable in the extreme tails of the distributions.

\subsection{A Phenomenological Langevin Equation}

At any given instant of time, there is a certain number of `buyers' which we
call $\phi_+(t)$ (the {\it demand}) and
`sellers', $\phi_-(t)$ (the {\it supply}). The first dynamical equation
describes the
effect of an offset between supply and demand, which tends to push the price
up (if
$\phi_+ > \phi_-$) or down in the other case. In general, one can write\footnote{For simplicity, we use in the
following a continuous time formalism, although discrete time evolution
equations would be more adapted.}:
\be
\frac{dx}{dt} = {\cal F}(\Delta \phi) \qquad \Delta \phi :=\phi_+ - \phi_-
\ee
where ${\cal F}$ is an increasing function, such that ${\cal F}(0)=0$. In the
following, we will frequently assume that ${\cal F}$ is linear (or else 
that $\Delta \phi$ is small enough to be satisfied with the first term in the
Taylor
expansion of ${\cal F}$), and write 
\be
\frac{dx}{dt} = \frac{\Delta \phi}{\lambda}
\ee
where $\lambda$ is a measure of {\it market depth} i.e. the excess demand
required to move the price by one unit.
When $\lambda$ is high, the market can 
`absorb' supply/demand offsets by very small price changes. Now, we try to
construct
a dynamical equation for the supply and demand separately. Consider for
example the
number of buyers $\phi_+$. Between $t$ and $t+dt$, a certain fraction of those
get
their deal and disappear (at least temporarily). This deal is usually ensured
by {\it
market makers}, which act as intermediaries between buyers and sellers. The
role of 
market markers is to absorb the demand (and supply) even if these do not match
perfectly. The effect of market makers
({\sc mm}) can thus be modelled as:
\be
\left.\frac{d\phi_\pm}{dt}\right|_{{\sc mm}} = -\Gamma_\pm \phi_\pm 
\ee
where $\Gamma_\pm$ are rates (inverse time scales). We furthermore assume that
market makers act
symmetrically, i.e, that $\Gamma_+=\Gamma_-=\gamma$. On liquid markets, the
time scale $1/\gamma$  before which a deal is reached is
short; typically a few minutes (see also below for another interpretation of
$1/\gamma$). 

There are several other effects which must be modeled to account for the time
evolution of supply and demand. One is the spontaneous ({\sc sp}) appearance
of new buyers
(or sellers), under the influence of new information, individual need for
cash, or particular investment strategies. This can be modelled as a white
noise term (not necessarily
Gaussian): 
\be
\left.\frac{d\phi_\pm}{dt}\right|_{{\sc sp}} = m_\pm(t) + \eta_\pm(t)
\ee
where $\eta_\pm$ have zero mean, and a short correlation time
$\tau_c$. $m_\pm$ is the average increase of demand (or supply), which might
also
depends on time through the time dependent anticipated return $R(t)$ and the
anticipated risk $\Sigma(t)$. It is quite clear that both these quantities are
constantly reestimated by the market participants, with a strong influence of
the
recent past. For example, `trend followers' extrapolate a local trend into the
future.
On the other hand, `fundamental analysts' estimate what they believe to be the
`true'
price of the stock; if the observed price is above this `true' price, the
anticipated
trend is reduced, and vice-versa. In mathematical terms, these effects can be
represented as: 
\be
R(t) =  R_0 + \alpha \int_{-\infty}^t dt' K_R(t-t') \frac{dx}{dt'} - \kappa
(x-x_0)
\ee
where $K_R$ is a certain kernel (of integral one) defining how the past
average trend is measured by the 
agents, and $\kappa$ is a mean-reversion force, towards the average (over the
fundamental analysts) `true
price' $x_0$ \footnote{Note that $x_0$ is actually itself time dependent,
although its
evolution in general takes place over rather long time scales (years).}. 

Similarly, the anticipated risk has a short time scale contribution. It is
well known
that an increase of volatility is badly felt by the agents, who immediately
increase
their estimate of risk. Hence, we write:
\be
\Sigma(t) =  \Sigma_0 + \beta \int_{-\infty}^t dt' K_\Sigma(t-t')
\left[\frac{dx}{dt'}\right]^2
\ee

Correspondingly, expanding $m_\pm(R,\Sigma)$ to lowest order, one has:
\be
m_\pm = m_{0\pm} +  \alpha_{\pm} \int_{-\infty}^t dt' K_R(t-t') \frac{dx}{dt'}
+ \beta_{\pm} \int_{-\infty}^t dt' K_\Sigma(t-t')
\left[\frac{dx}{dt'}\right]^2 - \kappa_\pm (x-x_0) \label{Eqmpm}
\ee
where the signs of the different coefficients are set by the observation that
$m_+$ is
an increasing function of return $R$ and a decreasing function of risk
$\Sigma$, and vice versa for $m_-$. Eq. (\ref{Eqmpm}) contains the leading
order
terms which arise if one assumes that the agents try to
reach a tradeoff between risk and return: the demand for an asset decreases
if is recent evolution shows high volatility and increases if it shows an
upward
trend. This is the case for example if the investors follow a
mean-variance optimisation
scheme with adaptive estimates of risk and return \cite{RefGen}, or the 
Black-Scholes option hedging strategy \cite{BS}. Actually, the 1987 crash is
often attributed (at least in part) to the automatic use of the 
Black-Scholes hedging strategy, which automatically generates sell orders when
the value of the stock goes down.

We are now in position to write an
equation for the supply/demand offset $\Delta \phi$ by summing all these
different
contributions:
\ba \nonumber
\frac{d \Delta \phi}{dt} &=& - \gamma \Delta \phi + m_0 
+ a \int_{-\infty}^t dt' K_R(t-t')
\frac{dx}{dt'} \cr
&-&  b \int_{-\infty}^t dt' K_\Sigma(t-t')
\left[\frac{dx}{dt'}\right]^2 - k (x-x_0) + \eta(t)\label{fund}
\ea
with $a,b,k > 0$. Note in particular that $b>0$ reflects the fact that agents
are
risk averse, and that an increase of the local volatility always leads to
negative
contribution to $\Delta \phi$. This feature will be crucial in the following.
For
definiteness, we will consider $\eta$ to be gaussian and normalize it as: 
 \be
\langle \eta(t) \eta(t') \rangle = 2\lambda^2
D \delta(t-t')
\ee 
where $D$ measure the susceptibility of the market to the random
external shocks, typically the arrival of information. As discussed in section
\ref{Empirical}, $D$
should also depend on the recent history, reflecting the fact that an 
increase in volatility induces a stronger reactivity of the market to external
news. In the same spirit as above, one could thus write:
\be
D = D_0 + D_1 \int_{-\infty}^t dt' K_D(t-t')
\left[\frac{dx}{dt'}\right]^2\label{d1}
\ee
For simplicity, we will neglect the influence of $D_1$ in the following. But
such a term might be responsible for the time reversal symmetry breaking
effect reported in section \ref{Empirical}, as well as for the fact that
volatility appears to be time dependent.

\subsection{Simple consequences. Liquid vs. Illiquid markets}

Let us consider the linear case `risk neutral' case where $b=0$. We will
assume
for simplicity that $K_R(t) = \Gamma \exp -(\Gamma t)$, and first consider the
local
limit where $\Gamma$ is much larger that $\gamma$ (short memory time). In this
case, the equation for
$x$  becomes that of an harmonic oscillator:
\be
\frac{d^2 x}{dt^2} + \left(\gamma-\frac{a}{\lambda}\right) \frac{dx}{dt} +
\frac{k}{\lambda} 
(x - \tilde x_0) = \frac{1}{\lambda} \eta(t) 
\ee
where $m_0$ has been absorbed into a redefinition of $\tilde x_0 := x_0 +
m_0/k$. For
liquid markets, where $\lambda$ and $\gamma$ are large enough, the `friction'
term $\tilde \gamma := \gamma-a/\lambda$ is positive. In this
case the market is stable, and the price oscillates around an equilibrium
value
$\tilde x_0$, which is higher than the average fundamental price if the
spontaneous
demand is larger than the spontaneous supply (i.e. $m_0$ is positive), as
expected when
the overall economy grows. The situation is rather different for illiquid
markets, or when trend following
effects are large, since $\tilde \gamma$ can be negative. In this case, the
market is
unstable, with an exponential rise or decay of the stock value, corresponding
to a
speculative bubble.  However, in this case, $dx/dt$ grows with time and it
soon
becomes untenable to neglect the higher order terms, in particular the risk
aversion
term proportional to $b$, which is responsible for a sudden collapse
\cite{Langevin}.  

\subsection{Risk aversion induced crashes}

In the case of liquid markets on short time scales, it is
reasonnable to set $k=0$ \cite{Langevin}. Setting $u=dx/dt$ and still focusing
on the limit where the memory time $\Gamma^{-1}$ is very
small, one finds the following non linear Langevin equation:
\be
\frac{du}{dt} = \frac{m_0}{\lambda} -\tilde \gamma u - \frac{b}{\lambda} u^2 +
\frac{1}{\lambda} \eta(t) \equiv -\frac{\partial V}{\partial u} 
+ \frac{1}{\lambda} \eta(t) \label{Langevin1}
\ee
This equation represents the evolution of the position $u$ of a viscous
fictitious particle
in a `potential' $V(u)$. In order to keep the mathematical form simple, we
set the average trend $m_0/\lambda$ to zero (no net average offset between
spontaneous demand and
spontaneous supply); this does not qualitatively change the following picture,
unless
$m_0$ is negative and large. The potential $V(u)$ can then be written as: 
\be V(u) =
\frac{\tilde \gamma}{2} u^2 + \frac{b}{3\lambda} u^3 \ee
which has a local minimum for $u=0$, and a local maximum for $u^* = - \lambda
\tilde
\gamma/b$, beyond which the potential plumets to $-\infty$. The `barrier
height' $V^*$
separating the stable region around $u=0$ from the unstable region is given
by:
\be
V^*=V(u^*)-V(0)= \frac{\tilde \gamma u^{*2}}{6}
\ee
The nature of the motion of $u$ in such a potential is the following: starting
at 
$u=0$, the particle has a random harmonic-like motion in the vicinity of $u=0$
until
an `activated' event (i.e. driven by the noise term) brings the particle near
$u^*$. Once this barrier is crossed,
the fictitious particle reaches $-\infty$ in finite time. In financial terms,
the
regime where $u$ oscillates around $u=0$ and where $b$ can be neglected, is
the
`normal' random walk regime discussed in the previous paragraph. (Note that
the random walk is biased when $m_0 \neq 0$). This normal regime
can however be interrupted by `crashes', where the time derivative of the
price
becomes very large and negative, due to the risk aversion term $b$ which
enhances the drop in the price. The point is that these two regimes can be
clearly separated since the average time $t^*$ needed for such crashes
to occur can be exponentially long, since it is given by the classical
Arrhenius-Kramers formula \cite{Wax,Hanggi}:
\be
t^* = 2 \pi \tau_1 \exp\left(\frac{V^*}{D}\right) = \frac{2 \pi}{\gamma} 
\exp\left(\frac{u^{*2}\tau_1}{3 \sigma^2}\right)\label{Arrh}
\ee
where $D$ is the variance of the noise $\eta$ and $\tau_1=1/\tilde \gamma$.
Taking $t^*=10$ years, $\sigma=1 \%$ per day, and
$\tau_1=10$ minutes, one finds that  the characteristic value $u^*$ beyond
which the
market `panics' and where a crash situation appears is of the order of $-1 \%$
in ten
minutes, which not unreasonnable. Note that in this line of thought, a crash
occurs because of an improbable succession
of unfavorable events, and not due to a single large event in particular.
Furthermore, there are
no `precursors' - characteristic patterns observed before the crash:
 before $u$ has reached $u^*$, it is impossible to decide whether it
will do so or whether it will quietly come back in the `normal' region $u
\simeq 0$. Note finally that an increase in the liquidity  factor 
$\gamma$ reduces the probability of crashes. This is related to the
stabilizing role of market makers, which appears very clearly.

The above calculations can be extended to the case where the `integration
time' of the market (appearing in the kernels $K_R$ and
$K_\Sigma$) is not very short, with very similar conclusions \cite{Langevin}. 

\section{Conclusion}

Research on financial markets can focus on rather different aspects. We have
discussed here three interrelated themes: the statistical nature of the
market fluctuations, its use for option pricing and the need for a
microsocopic `explicative' model which accounts for empirical 
observations. We
have emphasized the fact that a good model of these fluctuations was crucial
to estimate and control large risks. This requires in particular an adapted
theory for option pricing, which goes beyond the traditionnal Black-Scholes
dogma. We have shown that subtle statistical effects, such as the persistence
of volatility fluctuations, is rather well reflected in option prices. This
shows that the market as a whole behaves as an adaptive system, able to
correct (through trial and errors) the theoretical inadequacies of the 
Black-Scholes formula. 

The theoretical understanding of the {\it tails} of the distributions is
furthermore of fundamental importance, because by definition the empirical
observation of rare events leads does not allow one to determine the extreme
tails of the distribution with great accuracy. In this respect, an interesting
aspect of the Langevin equation discussed above is that crashes events appear
as fundamentally distinct from `normal' events, and their probability of
occurence is thus not expected to lie on the extrapolation of the (non
Gaussian) distribution
constructed from these `normal' events. This is actually what is observed:
although the empirical tail of the distribution of the S\&P 500's daily
increments is pretty well fitted by an exponential down to probabilities of
$10^{-3}$, events which have a probability of $10^{-4}$ (i.e. crashes which
occcur once every 40 years) have an amplitude which is much larger than
expected \cite{outliers}. 
 
There are many other aspects of financial markets which are worth
investigating using the tools and ideas of statistical physics, which we have
not discussed here at all: we refer the reader to the recent literature, which
can be found on the preprint server ``cond-mat''\cite{Rama}. One of the most
exciting subject is the
modelling of the interest curve, which might have a lot in common with the
dynamics of an elastic string driven by noise \cite{IRC}.

\subsection*{Acknowledgements}
This work is the result of many collaborations, which started with D. Sornette
and continued in a most fruitful and enjoyable 
way with Rama Cont and M. Potters. I also want to acknowledge 
important discussions with J.P. Aguilar, E. Aurell, P. Cizeau and 
L. Laloux.

\end{document}